\documentclass[aps, twocolumn, superscriptaddress, showpacs]{revtex4-1}
\pdfoutput=1
\usepackage{graphicx}
\usepackage{amssymb,amsmath}
\usepackage{bm}

\begin{document}

\title{Phase reduction and synchronization of a network of coupled dynamical elements exhibiting collective oscillations}

\author{Hiroya Nakao}
\email{nakao@mei.titech.ac.jp (corresponding author)}
\affiliation{Department of Systems and Control Engineering,
Tokyo Institute of Technology, Tokyo 152-8552, Japan}
\affiliation{Department of Mechanical Engineering, University of California, Santa Barbara, CA 93106, USA}

\author{Sho Yasui}
\affiliation{Department of Mechanical and Environmental Informatics,
Tokyo Institute of Technology, Tokyo 152-8552, Japan}

\author{Masashi Ota}
\affiliation{Department of Mechanical and Environmental Informatics,
Tokyo Institute of Technology, Tokyo 152-8552, Japan}
\affiliation{Ricoh Company Ltd., Japan}

\author{Kensuke Arai}
\affiliation{Department of Statistics and Mathematics, Boston University, Boston, MA 02215, USA}

\author{Yoji Kawamura}
\affiliation{Department of Mathematical Science and Advanced Technology,
Japan Agency for Marine-Earth Science and Technology, Yokohama 236-0001, Japan}

\begin{abstract}

A general phase reduction method for a network of coupled
dynamical elements exhibiting collective oscillations,
which is applicable to arbitrary networks of heterogeneous
dynamical elements, is developed.
A set of coupled adjoint equations for phase sensitivity
functions, which characterize phase response of the collective
oscillation to small perturbations applied to individual elements,
is derived.
Using the phase sensitivity functions, collective oscillation
of the network under weak perturbation can be 
described approximately by a one-dimensional phase equation.
As an example, mutual synchronization between a pair of collectively oscillating networks of excitable and oscillatory FitzHugh-Nagumo elements with random coupling is studied.

\end{abstract}

\maketitle

{\bf Networks of coupled dynamical elements exhibiting collective oscillations often play important functional roles in real-world systems. Here, a method for dimensionality reduction of such networks is proposed by extending the classical phase reduction method for nonlinear oscillators. By projecting the 
network state to a single phase variable,
a simple one-dimensional phase equation describing the collective oscillation is derived. As an example, synchronization between collectively oscillating random networks of neural oscillators is studied. The derived phase equation is general and will have wide applicability in control and optimization of collectively oscillating networks. }

\section{Introduction}

Synchronization of coupled dynamical elements is ubiquitously observed in the real world and often possesses important functional roles in biological and engineered systems~\cite{winfree80,pikovsky01,strogatz03}.
A series of beautiful experiments using finely tuned coupled electrochemical
oscillators by John L. Hudson and his collaborators~\cite{hudson0,hudson1,hudson2,hudson3,hudson4,hudson5,hudson6} has vividly revealed intriguing
synchronization dynamics that can occur in a network of coupled oscillators, including the first experimental realization of the collective synchronization transition, or Kuramoto transition, of globally coupled limit-cycle oscillators.

In the real world, it is often the case that a system is comprised of a number of different dynamical subsystems (elements), mutually coupled through an interaction network and exhibits stable collective oscillations, such as our body in which various organs mutually interact and obey the approximate 24h rhythm synchronized to the sun, or the power grids where synchronization of constituent AC generators is required for stable operation~\cite{motter13}.
A network of coupled chemical oscillators undergoing synchronized collective oscillations, intensively studied by Hudson~\cite{hudson0,hudson1,hudson2,hudson3,hudson4,hudson5,hudson6}, can be considered a fundamental experimental model of such collective dynamics.

In analyzing collective dynamics of a network of coupled dynamical elements,
one useful way is deriving a low-dimensional description of the collective
dynamics by reducing the dimensionality of the network.
For low-dimensional limit-cycle oscillators, the most successful and widely-used theoretical method for dimensionality reduction is the phase reduction~\cite{winfree80,kuramoto,hoppensteadt97,ermentrout10,brown04,ashwin16,nakao16}, where the dynamics of the oscillator is projected onto a single phase equation describing neutral dynamics along a one-dimensional stable limit cycle in the state space.

Generalization of the phase reduction method for high-dimensional systems exhibiting collective oscillations has recently been developed for coupled phase oscillators with global coupling~\cite{kawamura1} and with general network coupling~\cite{kori}, and for active rotators with global coupling~\cite{kawamura2}. Similar idea has been applied for the analysis of mutual synchronization between collectively oscillating populations of coupled phase oscillators~\cite{kawamura3,kawamura4,kawamura44}.
Moreover, the idea of collective phase reduction has further been generalized to spatially extended systems such as thermal convection~\cite{kawamura5,kawamura6} and reaction-diffusion systems exhibiting rhythmic spatio-temporal dynamics~\cite{nakao}.

In deriving a phase equation for the collective oscillation of a network, phase response of the network to external perturbations should be known. In Refs.~\cite{kawamura1,kori}, phase sensitivity functions for the collective oscillation are derived for a network of coupled phase oscillators.
However, the frameworks developed in Refs.~\cite{kawamura1,kori} are restrictive in that all the elements should be autonomous oscillators with approximately the same properties and their mutual coupling should be weak enough.
This hampers experimental investigation of phase response properties of collective oscillations in real-world systems, such as electrochemical oscillators.

In this paper, we extend the idea of collective phase reduction and derive a phase equation for a network of coupled dynamical elements in the most general form, where the dynamics of the elements can be arbitrary and the mutual interaction between the elements can be strong; the only assumption is that the whole network undergoes a stable collective limit-cycle oscillation.
We derive a set of coupled adjoint equations, which gives the phase sensitivity functions of the collective oscillation of the network to weak external perturbations applied to constituent dynamical elements, and reduce the dynamics of the whole network to a one-dimensional phase equation.
As an example, we calculate phase response property of a network of FitzHugh-Nagumo (FHN) elements exhibiting collective oscillations, where both excitable and oscillatory elements are coupled via random network connections,
and analyze mutual synchronization between a pair of FHN networks.

\begin{figure}
\center
\includegraphics[width=\hsize]{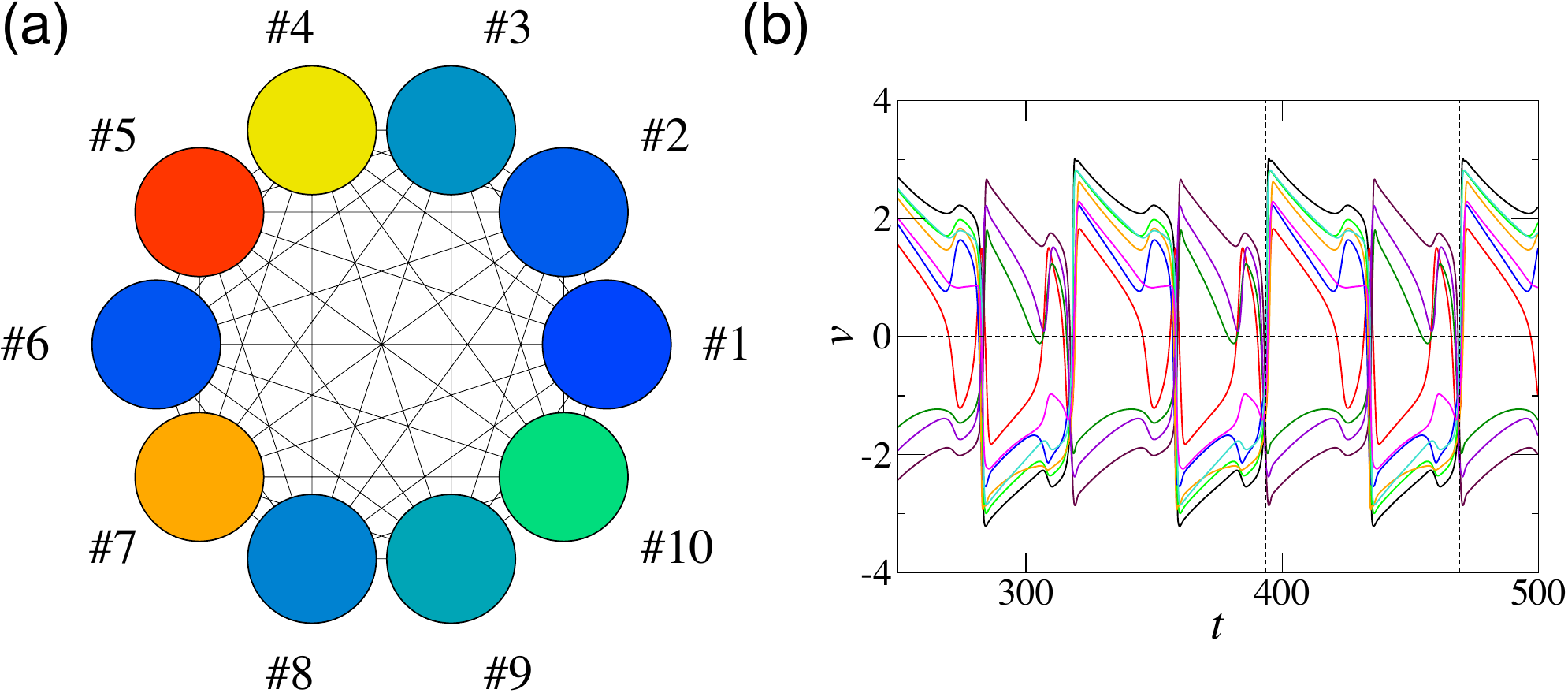}
\caption{(a) Schematic figure of a network of randomly coupled $10$ FHN elements (circles). Color reflects the variable $v$ (arbitrary scale). The numbers \#1-\#10 are indices of the elements (\#1-\#7: excitable, \#8-\#10: oscillatory). The color of each element corresponds to the value $v_i(t)$ ($i=1, 2, ..., 10$). (b) Collective oscillation of the network. Periodic dynamics of the $v$ components of the $10$ coupled FHN elements are shown. See Fig.~\ref{Fig1} for individual traces.}
\label{Fig0}
\end{figure}

\section{Phase reduction of a network of coupled dynamical elements}

\subsection{Phase reduction}

We consider a general network of $N$ coupled dynamical elements described by
\begin{align}
\frac{d}{dt} {\bm X}_i(t) = {\bm F}_i({\bm X}_i) + \sum_{j=1}^N {\bm G}_{ij} ( {\bm X}_i, {\bm X}_j )
\quad
(i=1, 2, ..., N), 
\label{1}
\end{align}
where ${\bm X}_i(t) \in {\bf R}^{m_i}$ is a $m_i \ (\geq 1)$-dimensional state of element $i$ at time $t$, ${\bm F}_i : {\bm R}^{m_i} \to {\bm R}^{m_i}$ represents individual dynamics of element $i$, and ${\bm G}_{ij} : {\bm R}^{m_i} \times {\bm R}^{m_j} \to {\bm R}^{m_i}$ describes the
effect of element $j$ on element $i$, respectively. It is assumed that ${\bm G}_{ii} = 0$ for all $i$, that is, self coupling does not exist or is absorbed into the individual part ${\bm F}_i$.
The dimensionality of each element does not need to be identical, and the dynamics ${\bm F}_i$ can differ from element to element. The interaction network ${\bm G}_{ij}$ can also be arbitrary as long as the network is connected and no element is isolated.

We assume that the whole network exhibits stable collective oscillation, i.e., the network possesses a stable limit-cycle solution
\begin{align}
{\bm X}_i^{(0)}(t+T) = {\bm X}_i^{(0)}(t) \quad (i=1, 2, ..., N)
\end{align}
of period $T$ and frequency $\omega = 2\pi / T$.
That is, each element repeats the same oscillatory behavior periodically with the same period $T$, though individual dynamics of the elements may differ from each other. See Fig.~\ref{Fig0} for an example.
We assume that such collective oscillation described by Eq.~(\ref{1}) is exponentially stable and persists even if subjected to weak perturbations.

Because the whole network exhibits collective oscillations, we can introduce a single collective phase variable $\theta(t) \in [0, 2\pi)$ of the network, which  increases with a constant natural frequency $\omega$ as
\begin{align}
\frac{d}{dt} \theta(t) = \omega,
\end{align}
and represent the state of the whole network (i.e., states of all the elements) as
\begin{align}
{\bm X}_i(t) = {\bm X}_i^{(0)}(\theta(t)) \quad (i=1, 2, ..., N)
\end{align}
as a function of the phase $\theta(t)$.

Now, suppose that the network described by Eq.~(\ref{1}), undergoing stable collective oscillations, is weakly perturbed as
\begin{align}
\frac{d}{dt} {\bm X}_i(t) = {\bm F}_i({\bm X}_i) + \sum_{j=1}^N {\bm G}_{ij} ( {\bm X}_i, {\bm X}_j ) + \epsilon {\bm p}_i(t)
\cr
(i=1, 2, ..., N),
\label{2}
\end{align}
where ${\bm p}_i(t) \in {\bm R}^{m_i}$ represents the external perturbation given to the element $i$ at time $t$ and $0 < \epsilon \ll 1$ is a small parameter representing the intensity of the perturbation.
Because the whole network can be seen as a single big limit-cycle oscillator, by generalizing the standard phase reduction method~\cite{kuramoto}, we can approximately represent the dynamics of the whole network by using a single scalar equation for the collective phase $\theta(t)$ when $\epsilon$ is sufficiently small.

As derived in Appendix, the approximate phase equation for the collective phase $\theta(t)$, which is correct up to $O(\epsilon)$, is given by
\begin{align}
\frac{d}{dt}{\theta}(t) = \omega + \epsilon \sum_{i=1}^N {\bm Q}_i(\theta) \cdot {\bm p}_i(t),
\label{phaseeq}
\end{align}
where ${\bm Q}_i(\theta) \in {\bm R}^{m_i}$ is the phase sensitivity function of the element $i$ ($i=1, 2, ..., N$).
Thus, we can individually evaluate the effect of external perturbation ${\bm p}_i(t)$ applied to each element $i$ on the phase $\theta(t)$ of the collective oscillation of the network, and approximately describe the collective oscillation of the whole network by a simple reduced phase equation.

As derived in Appendix, the phase sensitivity functions ${\bm Q}_i(\theta)$ are given by a $2\pi$-periodic solution to the following set of coupled adjoint equations,
\begin{align}
\omega \frac{d}{d\theta} {\bm Q}_i(\theta) = - {\rm J}_i^{\dag}(\theta) {\bm Q}_i(\theta)
- \sum_{j=1}^N {\rm M}_{ij}^{\dag}(\theta) {\bm Q}_i(\theta)
\cr
- \sum_{j=1}^N {\rm N}^{\dag}_{ji}( \theta ) {\bm Q}_j(\theta)
\quad
(i=1, 2, ..,. N),
\label{adjoint}
\end{align}
where ${\rm J}_i(\theta) = \partial{\bm F}_i({\bm X}_i) / \partial {\bm X}_i \in {\bm R}^{m_i \times m_i}$, ${\rm M}_{ij}(\theta) = \partial {\bm G}_{ij}({\bm X}_i, {\bm X}_j) / \partial {\bm X}_i \in {\bm R}^{m_i \times m_i}$, and ${\rm N}_{ij}(\theta) = \partial {\bm G}_{ij}({\bm X}_i, {\bm X}_j) / \partial {\bm X}_j\in {\bm R}^{m_i \times m_j}$ are Jacobian matrices of ${\bm F}_i$ and ${\bm G}_{ij}$ evaluated at ${\bm X}_i = {\bm X}_i^{(0)}(\theta)$, respectively, and
$\dag$ indicates matrix transpose.
Also, the phase sensitivity functions should satisfy the normalization condition
\begin{align}
\sum_{i=1}^N {\bm Q}_i(\theta) \cdot \frac{d {\bm X}_i^{(0)}(\theta)}{d\theta} = 1.
\label{normalization}
\end{align}
By numerically finding a $2\pi$-periodic solution to the adjoint equation (\ref{adjoint}) with the normalization condition (\ref{normalization}),
we can obtain the phase sensitivity functions ${\bm Q}_i(\theta)$ and evaluate the effect of weak perturbations  given to the dynamical elements on the collective phase.

Note that the above result is applicable to arbitrary networks of coupled dynamical elements, where coupling networks and properties of constituent elements are arbitrary. The only assumption is that the whole network has a stable limit-cycle solution.
When the network under consideration is of a reaction-diffusion type, the above results can be related to the previous results on continuous reaction-diffusion media (see Appendix).

\subsection{Synchronization between a pair of interacting networks}

A representative application of the reduced phase equation is the analysis of synchronization properties of mutually coupled oscillating networks. We here consider mutual synchronization between a pair of symmetrically coupled networks
with identical properties, A and B, given by
\begin{align}
\frac{d}{dt} {\bm X}^A_i(t) = {\bm F}_i({\bm X}^A_i) &+ \sum_{j=1}^N {\bm G}_{ij} ( {\bm X}^A_i, {\bm X}^A_j )
\cr
&+ \epsilon \sum_{j=1}^N {\bm H}_{ij} ( {\bm X}_i^A, {\bm X}_j^B ),
\cr
\frac{d}{dt} {\bm X}^B_i(t) = {\bm F}_i({\bm X}^B_i) &+ \sum_{j=1}^N {\bm G}_{ij}  ( {\bm X}^B_i, {\bm X}^B_j )
\cr
&+ \epsilon \sum_{j=1}^N {\bm H}_{ij} ( {\bm X}_i^B,  {\bm X}_j^A ),
\label{twocoupled}
\end{align}
where ${\bm X}^A_i$ and ${\bm X}^B_i$ are the state variables of elements $i=1, ..., N$ in networks $A$ and $B$, respectively, ${\bm H}_{ij}({\bm X}_i^A, {\bm X}_j^B)$ represents inter-network coupling between ${\bm X}^A_i$ and ${\bm X}^B_j$,
and $\epsilon$ is a small parameter.
For simplicity, it is assumed that the two networks are identical, i.e., they share the same parameter values for the elements and the same internal coupling network ${\bm G}_{ij}$.
It is also assumed that collective oscillation of each network persists when small mutual interaction between the networks is introduced.

We denote the collective phase of the two networks as $\theta^A(t)$ and $\theta^B(t)$, respectively. Then, by using  the phase sensitivity functions ${\bm Q}_i$, which are common to both networks, the dynamics of the above two-coupled networks can be reduced to a pair of coupled phase equations, which is correct up to $O(\epsilon)$, as
\begin{align}
\frac{d}{dt} \theta^A(t) &= \omega + \epsilon \sum_{i=1}^N {\bm Q}_i(\theta^A) \cdot \sum_{j=1}^N {\bm H}_{ij} ( {\bm X}_i^{(0)}(\theta^A),  {\bm X}_j^{(0)}(\theta^B) ),
\cr
\frac{d}{dt} \theta^B(t) &= \omega + \epsilon \sum_{i=1}^N {\bm Q}_i(\theta^B) \cdot \sum_{j=1}^N {\bm H}_{ij} ( {\bm X}_i^{(0)}(\theta^B),  {\bm X}_j^{(0)}(\theta^A) ).
\end{align}
Now, by following the standard procedure of phase reduction theory~\cite{kuramoto,hoppensteadt97,ermentrout10} and invoking averaging approximation, these equations can be transformed to
\begin{align}
\frac{d}{dt} \theta^A(t) &= \omega + \epsilon \sum_{i=1}^N \sum_{j=1}^N \Gamma_{ij}( \theta^A - \theta^B ),
\cr
\frac{d}{dt} \theta^B(t) &= \omega + \epsilon \sum_{i=1}^N \sum_{j=1}^N \Gamma_{ij}( \theta^B - \theta^A ),
\label{twocplphase}
\end{align}
which is also correct up to $O(\epsilon)$, where
\begin{align}
\Gamma_{ij}(\phi) &= \frac{1}{2\pi} \int_0^{2\pi} d\psi
\cr
&\times
{\bm Q}_i(\psi+\phi) \cdot {\bm H}_{ij} ( {\bm X}_i^{(0)}(\psi+\phi), {\bm X}_j^{(0)}(\psi) )
\quad \quad
\end{align}
is the phase coupling function between the elements $i$ and $j$ of the two networks.
From Eq.~(\ref{twocplphase}), the phase difference $\phi = \theta^A - \theta^B$ between the networks obeys
\begin{align}
\frac{d}{dt} \phi(t) 
= \epsilon \Gamma_a(\phi),
\label{phasedifeq}
\end{align}
where
\begin{align}
\Gamma_a(\phi) = \sum_{i=1}^N \sum_{j=1}^N [ \Gamma_{ij}(\phi) - \Gamma_{ij}(-\phi) ],
\label{gamma}
\end{align}
is an antisymmetric function of $\phi$, i.e., $\Gamma_a(\phi) = - \Gamma_a(-\phi)$.
Thus, by calculating $\Gamma_a(\phi)$, we can predict the stable phase differences between the two networks as the stable fixed point of the one-dimensional phase equation~(\ref{phasedifeq}).

\begin{figure}
\center
\includegraphics[width=\hsize]{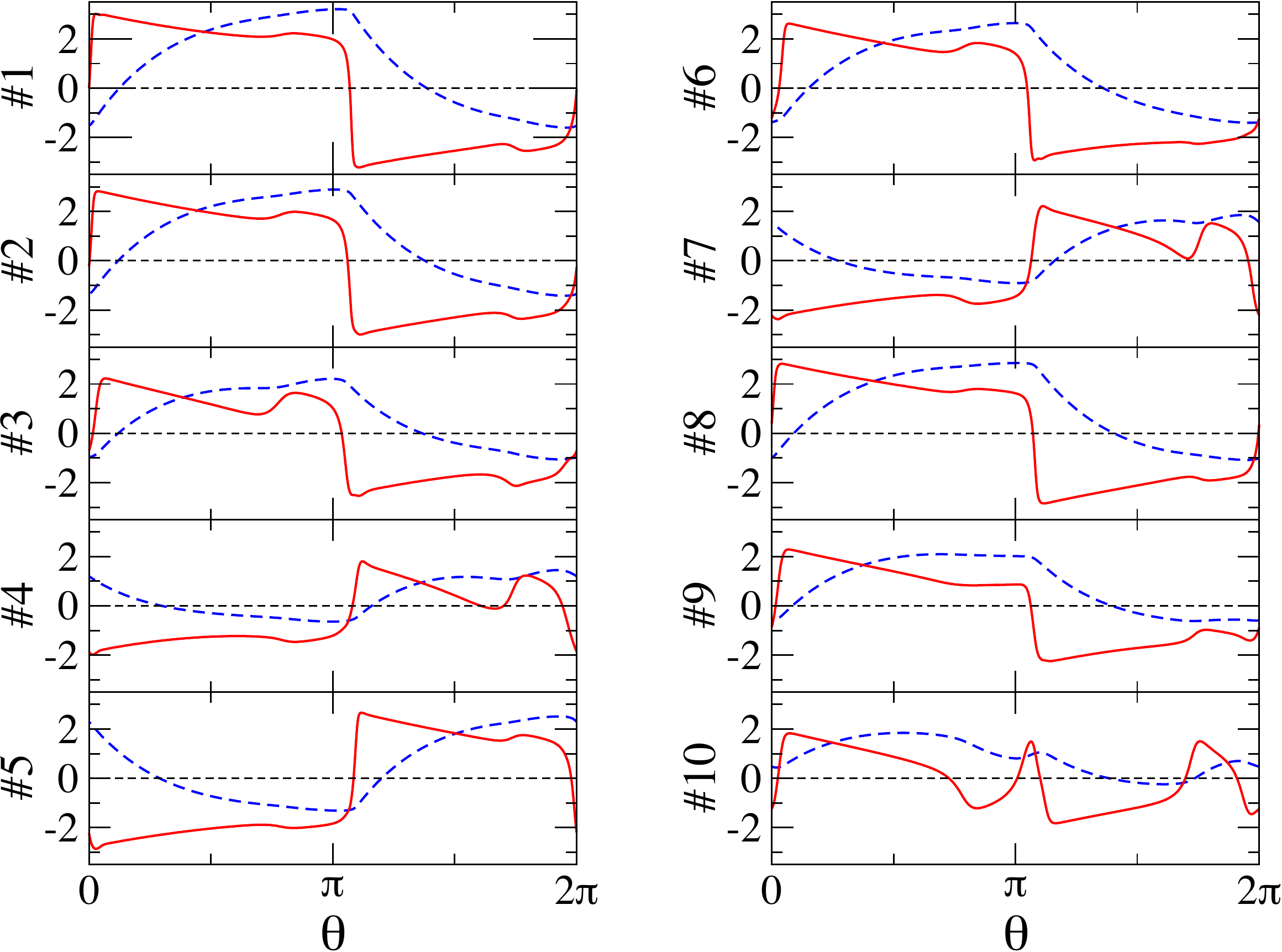}
\caption{Dynamics of $u$ and $v$ components of coupled FHN elements for one period of oscillation. Each figure shows $u_i^{(0)}(\theta)$ (blue dashed) and $v_i^{(0)}(\theta)$ (red solid) of $i$th element ($i=1, 2, ..., 10$) in the steadily oscillating state.}
\label{Fig1}
\end{figure}

\begin{figure}
\center
\includegraphics[width=\hsize]{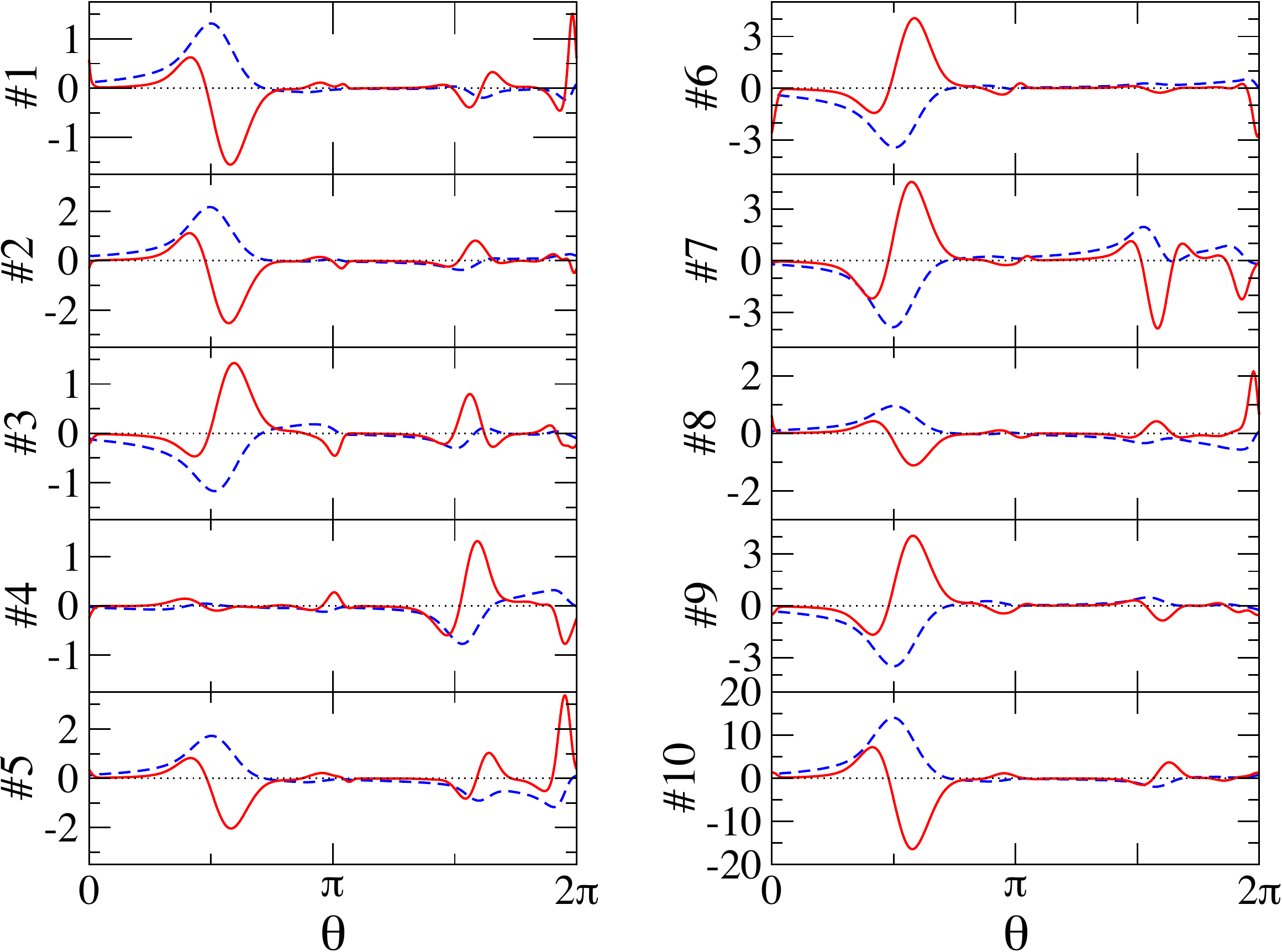}
\caption{Phase sensitivity functions of the network of FHN elements. Each figure shows $Q^u_i(\theta)$ (blue dashed) and $Q^v_i(\theta) \times 5$ (red solid) of $i$th element ($i=1, 2, ..., 10$), where $Q^v_i(\theta)$ is multiplied by $5$ for visual clarity.}
\label{Fig2}
\end{figure}

\section{Example}

\subsection{A network of coupled FitzHugh-Nagumo elements}

As an example, we consider a network of $N$ coupled FitzHugh-Nagumo elements~\cite{hoppensteadt97,ermentrout10} with random connections. 
The state variable of each element $i$ ($i=1, 2, ..., N$) is two-dimensional,
\begin{align}
{\bm X}_i = (u_i, v_i)^{\dag},
\end{align}
which obeys 
\begin{align}
{\bm F}_i({\bm X}_i) =
\left(\delta (a + v_i - b u_i), \ v_i - \frac{v_i^3}{3} - u_i + I_i\right)^{\dag}.
\end{align}
We assume that the parameter $I_i$ of the element can differ
between the elements, so the elements can be either oscillatory
or excitable
depending on $I_i$. The other parameters are assumed to be identical.
We also assume that only the $v$ component (which is related to the
membrane potential of a neuron) can diffuse over the network
and the mutual coupling between elements $i$ and $j$ is given by
\begin{align}
{\bm G}_{ij}({\bm X}_i, {\bm X}_j) = K_{ij}
(0, \  v_j - v_i )^{\dag},
\label{fhncoupling}
\end{align}
where $K_{ij} \in {\bm R}$ is the $(i, j)$ component of an ${N \times N}$ matrix ${\rm K}$ representing the coupling network.

In the numerical simulations, we consider $N=10$ FitzHugh-Nagumo elements. The parameters of the elements are $I_i = 0.2$ for the elements $i=1, ..., 7$, which exhibit excitable dynamics, and $I_i=0.8$ for the elements $i=8, ..., 10$, which exhibit self-oscillatory dynamics. The other parameters are $\delta = 0.08$, $a = 0.7$, and $b = 0.8$.
Each component $K_{ij}$ of the coupling matrix ${\rm K}$ is randomly and independently drawn from a uniform distribution $[-0.6, 0.6]$.
The initial conditions of the elements are taken to be $u_i = 1$ and $v_i = 1$ for all $i$. See Appendix C for the actual ${\rm K}$ used in the simulations and a brief description of the qualitative dynamics of the network.
Note that the coupling matrix is not symmetric and each component can take both positive and negative values, so some pairs of the elements are mutually attractive while some other pairs are repulsive with differing coupling intensities.

With these parameter values, the whole network exhibits a limit-cycle oscillation in the $20$-dimensional state space of period $T \simeq 75.73$, where each element $i=1, ..., 10$ repeats its own dynamics periodically.
Figure~\ref{Fig0} schematically shows a network of $10$ coupled FHN elements and an example of the limit-cycle oscillation of the whole network, where $v$ components of the FHN elements are shown.
It can be seen that the dynamics of the elements are different from each other because of the heterogeneity of the elements and the random network connections between them, but the whole dynamics exhibits collective oscillation of period $T$.
We denote this limit-cycle solution as
\begin{align}
{\bm X}^{(0)}_i(\theta) = ( u_i^{(0)}(\theta), v_i^{(0)}(\theta) )^{\dag}
\quad (i=1, 2, ..., 10)
\end{align}
as function of the phase $0 \leq \theta < 2\pi$.
Figure~\ref{Fig1} shows the dynamics of the $u$ and $v$ components of the elements $i=1, ..., 10$ for one period of collective oscillation as a function of $\theta$, showing mutually similar but different oscillatory dynamics.
It can be confirmed numerically that this collective oscillation is stable and persists even if perturbed by weak external disturbances.

\begin{figure}
\center
\includegraphics[width=\hsize]{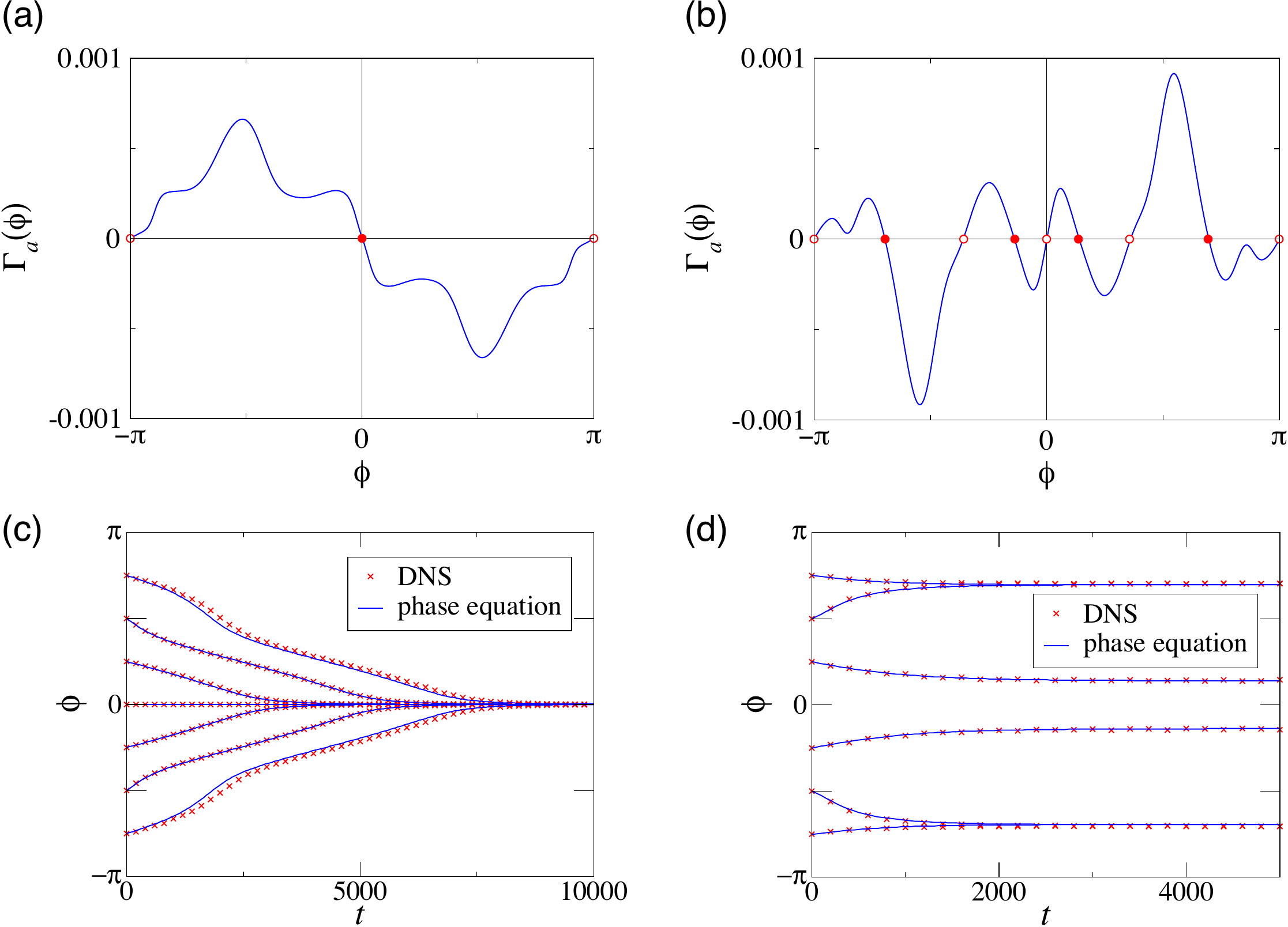}
\caption{Antisymmetric part of the phase coupling functions (a), (b) and dynamics of the phase differences (c), (d). In (a) and (c), results for the Case 1 [Eq.~(\ref{case1})] are shown, and in (b) and (d), results for the Case 2 [Eq.~(\ref{case2})] are shown. In (c) and (d), evolution of the phase differences starting from several initial values are shown, where results obtained by direct numerical simulations (DNS) of the FHN networks are compared with those obtained by the reduced phase equation.   }
\label{Fig3}
\end{figure}

\begin{figure}
\center
\includegraphics[width=\hsize]{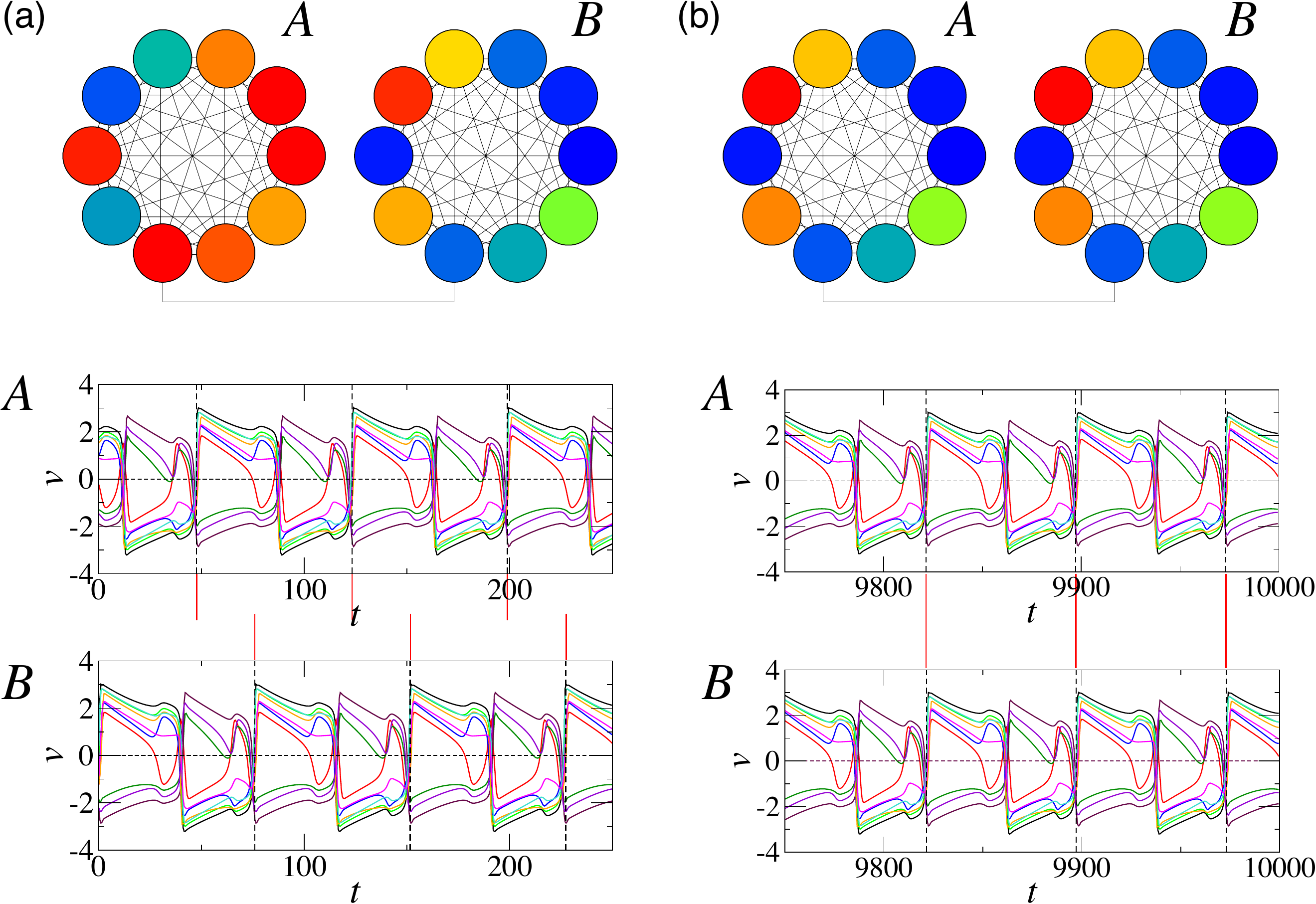}
\caption{In-phase synchronization of collective oscillations of the two FHN networks for the Case 1. (a) initial states and (b) synchronized states sufficiently after relaxation. The networks shown in the top panels of (a) and (b) are plotted in the same way as in Fig.~1; see Fig.~1 for the indices of the elements. The color of each element in the networks corresponds to the value $v^A_i(t)$ or $v^B_i(t)$ ($i=1, 2, ..., 10$). }
\label{Fig4}
\end{figure}

\subsection{Phase sensitivity functions}

The Jacobian matrices of ${\bm F}_i$ and ${\bm G}_{ij}$ are given by
\begin{align}
{\rm J}_i(\theta) = \begin{pmatrix} - \delta b & \delta \\ -1 & 1 - \{ v^{(0)}_i(\theta) \}^2 \end{pmatrix},
\end{align}
\begin{align}
{\rm M}_{ij} = K_{ij} \begin{pmatrix} 0 & 0 \\ 0 & -1 \end{pmatrix},
\quad
{\rm N}_{ij} = K_{ij} \begin{pmatrix} 0 & 0 \\ 0 & 1 \end{pmatrix},
\end{align}
for $i \neq j$, and ${\rm M}_{ij} = 0$, ${\rm N}_{ij} = 0$ when $i=j$.
By numerically solving the adjoint equations~(\ref{adjoint}) with these Jacobian matrices, we obtain the phase sensitivity functions
\begin{align}
{\bm Q}_i(\theta) = ( Q^u_i(\theta),\ Q^v_i(\theta) )^{\dag}
\quad (i=1, 2, ..., 10)
\end{align}
as their $2\pi$-periodic solutions.

Figure~\ref{Fig2} shows the phase sensitivity functions ${\bm Q}_i(\theta)$ of all elements $i=1, 2, ..., 10$. The phase sensitivity functions are different from element to element, again reflecting the heterogeneity and random coupling of the elements.
In this particular example, the phase sensitivity function of the $10$th element, which exhibits qualitatively different dynamics from other elements in Fig.~\ref{Fig1} due to relatively strong coupling, has considerably larger amplitudes than those of the other elements.

\subsection{Synchronization between a pair of FitzHugh-Nagumo networks}

We now analyze phase synchronization between
a pair of symmetrically coupled identical networks
described by Eq.~(\ref{twocoupled}) with $N=10$ FitzHugh-Nagumo elements.
Each network is as described in the previous subsections, and the inter-network
coupling is assumed to be
\begin{align}
{\bm H}_{ij}({\bm X}^A_i, {\bm X}^B_j) = C_{i,j} (0, \  v^B_j - v^A_i )^{\dag},
\end{align}
where again only the $v$ components are coupled between the networks $A$ and $B$, and the matrix $C_{i, j} \in {\bm R}^{N \times N}$ determines if the elements $i$ in network $A$ and $j$ in network $B$ are connected.
The small parameter determining the intensity of mutual coupling is fixed at $\epsilon=0.005$.

As an example, we consider two types of the inter-network coupling matrices $C_{i,j}$,
\begin{align}
&\mbox{Case 1} : \quad C_{8,8} = 1, \quad C_{i,j} = 0 \quad \mbox{(otherwise)},
\label{case1}
\\
&\mbox{Case 2} : \quad C_{2,10} = C_{5,7} = 1, \quad C_{i,j} = 0 \quad \mbox{(otherwise)}.
\label{case2}
\end{align}
For each case, the antisymmetric part $\Gamma_a(\phi)$ of the phase coupling function is shown in Figs.~\ref{Fig3}(a) and (b).
From Eq.~(\ref{phasedifeq}) for the phase difference $\phi$, we can identify the stable phase differences between the networks as the zero-crossing points of $\Gamma_a(\phi)$ with negative slopes.
Depending on $C_{i,j}$, it is predicted that the two networks undergo in-phase synchronization with zero phase difference (Case 1), or converge to either of four stable phase differences (Case 2) depending on the initial condition.

To confirm the prediction of the reduced phase equation, we numerically calculate the evolution of the phase differences between the two FHN networks by direct numerical simulations and compare them with those obtained from the reduced phase equations
in Figs.~\ref{Fig3}(c) and (d).
From the figures, we see that the two networks indeed synchronize at the stable phase differences predicted by the phase equations, as illustrated in Fig~\ref{Fig4}.

\section{Discussion}

We have formulated a phase reduction framework for a network of coupled dynamical
networks exhibiting collective oscillations.
Though we have treated only a simple example of two identical networks of neural oscillators, the theory is general and can be applied to analyzing, controlling, and designing networks of dynamical elements exhibiting collective oscillations.
Several interesting directions would be optimization of injection locking of the collective oscillation of a network~\cite{moehlis06,harada10,zlotnik13,pikovsky15,tanaka15,zlotnik16}, optimization of mutual coupling between the networks for synchronization~\cite{shirasaka17,kawamura17,kawamura16}, and design of network structures that lead to desirable phase response properties.
Because the theory does not require homogeneity of the dynamical elements
nor smallness of the coupling of the network, the theory can be tested by real experimental systems, such as the system of coupled electrochemical oscillators developed by John L. Hudson.

\begin{acknowledgments}

The idea of the present study originated from past discussion with H.~Kori, I.~Z.~Kiss, C.~G.~Rusin, Y.~Kuramoto, and J.~L.~Hudson. Their useful comments are gratefully acknowledged.

A decade ago, one of the authors (H.N.) shared an office with Prof. Hudson (Jack) during their stay at Prof. Mikhailov's group at Fritz-Haber Institute in Berlin. H.N. is deeply indebted to Jack's friendly advice. An elder, experienced professor's words can be precious to a young researcher and influence his scientific career. H.N. would like to dedicate this study to the memory of Jack.

H.N. acknowledges financial support from JSPS (Japan) KAKENHI Grant Numbers JP16H01538, JP16K13847, and JP17H03279.
Y.K. acknowledges financial support from JSPS (Japan) KAKENHI Grant Number JP16K17769.

\end{acknowledgments}

\section*{Appendix}

\subsection{Derivation of adjoint equations}

We here derive the adjoint equations for the phase sensitivity functions by generalizing the argument in Ref.~\cite{nakao}.
We assume that the network possesses a stable limit-cycle solution ${\bm X}_i^{(0)}(t)$ of period $T$ in the $(\sum_{i=1}^N m_i)$-dimensional state space, and initial states of the network around this limit cycle are exponentially attracted to this limit cycle.
We first define a phase function of the network,
\begin{align}
\theta = \Theta({\bm X}_1,&{\bm X}_2, ..., {\bm X}_N): 
\cr
&{\bm R}^{m_1} \times {\bm R}^{m_2} \times ... \times {\bm R}^{m_N} \to [0, 2\pi),
\end{align}
which increases with a constant frequency $\omega = 2 \pi / T$ in the whole basin of attraction of the limit cycle.
That is, we require that
\begin{align}
\frac{d}{dt} \theta(t)
=& \sum_{i=1}^N \frac{\partial \Theta}{\partial {\bm X}_i} \cdot \frac{d {\bm X}_i}{dt}
\cr
=& \sum_{i=1}^N \frac{\partial \Theta}{\partial {\bm X}_i}
\cdot \left( {\bm F}_i({\bm X}_i) + \sum_{j=1}^N {\bm G}_{ij}({\bm X}_i, {\bm X}_j) \right) = \omega,
\cr
\end{align}
where $\partial \Theta / \partial {\bm X}_i$ represents the gradient of $\Theta$ with respect to the variable ${\bm X}_i$.
If the network is perturbed as in Eq.~(3), the phase obeys
\begin{align}
\frac{d}{dt} \theta(t) &= \sum_{i=1}^N \frac{\partial \Theta}{\partial {\bm X}_i} \cdot \left( {\bm F}_i({\bm X}_i) + \sum_{j=1}^N {\bm G}_{ij}({\bm X}_i, {\bm X}_j) + \epsilon {\bm p}_i(t) \right)
\cr
&= \omega + 
\epsilon \sum_{i=1}^N \frac{\partial \Theta}{\partial {\bm X}_i} \cdot {\bm p}_i(t),
\end{align}
which is not yet closed in $\theta$ because the gradient terms depend on all ${\bm X}_i$.
To close the equation, we consider the case that the perturbation is sufficiently small, that is, $0 < \epsilon \ll 1$, and the state of the network stays in the vicinity of the limit cycle,
\begin{align}
{\bm X}_i(t) = {\bm X}^{(0)}_i(\theta(t)) + O(\epsilon).
\end{align}
Then, the gradient term can be approximated on the limit-cycle solution as
\begin{align}
{\bm Q}_i(\theta) = \left. \frac{\partial \Theta}{\partial {\bm X}_i} \right|_{\{{\bm X}_i = {\bm X}_i^{(0)}(\theta)\}_{i=1, 2, ..., N}},
\end{align}
and we can obtain an approximate phase equation that is closed in $\theta$ as
\begin{align}
\frac{d}{dt}{\theta}(t) = \omega + \epsilon \sum_{i=1}^N {\bm Q}_i(\theta(t)) \cdot {\bm p}_i(t) + O(\epsilon^2).
\end{align}
We call ${\bm Q}_i(\theta)$ the phase sensitivity function of element $i$.

It is of course difficult to explicitly obtain the phase function $\Theta$ for general networks, but we can derive a set of equations (adjoint equations) that determine ${\bm Q}_i(\theta)$ by extending the elegant derivation by Brown, Moehlis, and Holmes~\cite{brown04}.
Suppose a network state on the limit cycle, $\{ {\bm X}^{(0)}_1(\theta),\ ...,\ {\bm X}^{(0)}_N(\theta) \}$, and another network state close to it, $\{ {\bm X}_1 = {\bm X}^{(0)}_1(\theta) + \epsilon {\bm y}_1,\ ...,\ {\bm X}_N = {\bm X}^{(0)}_N(\theta) + \epsilon {\bm y}_N \}$, where $\epsilon {\bm y}_i \in {\bm R}^{m_i} \ (i=1, ..., N)$ represent small variations.
We represent the phase of the first state as
\begin{align}
\theta = \Theta({\bm X}^{(0)}_1(\theta),\ ...,\ {\bm X}^{(0)}_N(\theta)), 
\label{phase1}
\end{align}
and that of the second state as
\begin{align}
\theta'
=& \Theta({\bm X}_1,\ ...,\ {\bm X}_N)
\cr
=& \Theta({\bm X}^{(0)}_1(\theta) + \epsilon {\bm y}_1,\ ...,\ {\bm X}^{(0)}_N(\theta)+ \epsilon {\bm y}_N).
\end{align}
By the definition of the phase function, the difference $\Delta \theta(t) = \theta'(t) - \theta(t)$ remains constant when the perturbation is absent, because both $\theta(t)$ and $\theta'(t)$ increase with the same frequency $\omega$.
When the variations are sufficiently small, the difference between these two phases can be represented as
\begin{align}
\Delta \theta =& \Theta({\bm X}^{(0)}_1(\theta) + \epsilon {\bm y}_1, ..., {\bm X}^{(0)}_N(\theta) + \epsilon {\bm y}_N)
\cr
&- \Theta({\bm X}^{(0)}_1(\theta), ..., {\bm X}^{(0)}_N(\theta))
\cr
=& \Theta({\bm X}^{(0)}_1(\theta), ..., {\bm X}^{(0)}_N(\theta))
+ \epsilon \sum_{i=1}^N \left. \frac{\partial \Theta}{\partial {\bm X}_i}\right|_{\{ {\bm X}_i = {\bm X}^{(0)}_i(\theta) \} } \cdot {\bm y}_i
\cr
&- \Theta({\bm X}^{(0)}_1(\theta), ..., {\bm X}^{(0)}_N(\theta))
 + O(\epsilon^2)
\cr
=& \epsilon \sum_{i=1}^N \left. \frac{\partial \Theta}{\partial {\bm X}_i}\right|_{\{ {\bm X}_i = {\bm X}^{(0)}_i(\theta) \} } \cdot {\bm y}_i + O(\epsilon^2)
\cr
=& \epsilon \sum_{i=1}^N {\bm Q}_i(\theta) \cdot {\bm y_i} + O(\epsilon^2),
\end{align}
where we assumed that the phase function can be expanded in Taylor series.
Thus, the phase difference should satisfy
\begin{align}
\frac{d}{dt} \Delta \theta(t) = \epsilon \sum_{i=1}^N
\left( \frac{d{\bm Q}_i(\theta)}{dt} \cdot {\bm y}_i + {\bm Q}_i(\theta) \cdot \frac{d{\bm y}_i}{dt} \right) = 0
\label{phasedif}
\end{align}
at the first order approximation in $\epsilon$.

Now, from Eq.~(\ref{1}), the variations ${\bm y}_i(t)$ obey linearized equations
\begin{align}
\frac{d}{dt} {\bm y}_i(t) = {\rm J}_i(\theta(t)) {\bm y}_i(t)
&+ \sum_{j=1}^{N} {\rm M}_{ij}(\theta(t)) {\bm y}_i(t)
\cr
&+ \sum_{j=1}^{N} {\rm N}_{ij}(\theta(t)) {\bm y}_j(t)
\label{variational}
\end{align}
for $i=1, 2, ..., N$, where
\begin{align}
&{\rm J}_i(\theta) = \left. \frac{\partial {\bm F}_i({\bm X})}{\partial {\bm X}} \right|_{{\bm X} = {\bm X}_i^{(0)}(\theta)} \in {\bm R}^{m_i \times m_i},
\\
&{\rm M}_{ij}(\theta) = \left. \frac{\partial {\bm G}_{ij}({\bm X}, {\bm Y})}{\partial {\bm X}} \right|_{{\bm X} = {\bm X}_i^{(0)}(\theta),\ {\bm Y} = {\bm X}_j^{(0)}(\theta)} \in {\bm R}^{m_i \times m_i},
\end{align}
and
\begin{align}
{\rm N}_{ij}(\theta) =  \left. \frac{\partial {\bm G}_{ij}({\bm X}, {\bm Y})}{\partial {\bm Y}} \right|_{{\bm X} = {\bm X}_i^{(0)}(\theta),\ {\bm Y} = {\bm X}_j^{(0)}(\theta)} \in {\bm R}^{m_i \times m_j}
\end{align}
are the Jacobian matrices of ${\bm F}_i$ and ${\bm G}_{ij}$. Note that ${\rm J}_i$ and ${\rm M}_{ij}$ are $m_i \times m_i$ square matrices, while ${\rm N}_{ij}$ is generally a non-square matrix, and ${\rm N}_{ii}$ and ${\rm M}_{ii}$ are zero matrices because ${\bm G}_{ii} = 0$ for all $i$.
Plugging Eq.~(\ref{variational}) into Eq.~(\ref{phasedif}), we obtain
\begin{align}
0
=& \sum_{i=1}^N \Big( \frac{d{\bm Q}_i(\theta)}{dt} \cdot {\bm y}_i + {\bm Q}_i(\theta) \cdot \Big[ {\rm J}_i(\theta) {\bm y}_i
\cr
&+ \sum_{j=1}^{N} {\rm M}_{ij}(\theta) {\bm y}_i + \sum_{j=1}^{N} {\rm N}_{ij}(\theta) {\bm y}_j \Big] \Big)
\cr
=& \sum_{i=1}^N \Huge( \omega \frac{d{\bm Q}_i(\theta)}{d\theta} \cdot {\bm y}_i +  {\rm J}_i(\theta)^{\dag} {\bm Q}_i(\theta) \cdot {\bm y}_i
\cr
&+ \sum_{j=1}^{N} {\rm M}_{ij}^{\dag} {\bm Q}_i(\theta) \cdot {\bm y}_i + \sum_{j=1}^{N} {\rm N}_{ij}^{\dag}(\theta) {\bm Q}_i(\theta) \cdot {\bm y}_j \Big),
\ \ \ 
\label{middle}
\end{align}
where $\dag$ indicates matrix transpose and $d\theta/dt = \omega$ is used. By rewriting the last term as
\begin{align}
\sum_{i=1}^N \sum_{j=1}^N {\rm N}_{ij}^{\dag}(\theta) {\bm Q}_i(\theta) \cdot {\bm y}_j
=
\sum_{i=1}^N \sum_{j=1}^N {\rm N}_{ji}^{\dag}(\theta) {\bm Q}_j(\theta) \cdot {\bm y}_i,
\end{align}
we can further transform Eq.~(\ref{middle}) as
\begin{align}
&
\sum_{i=1}^N \Big( \omega \frac{d{\bm Q}_i(\theta)}{d\theta} + {\rm J}_i(\theta)^{\dag} {\bm Q}_i(\theta)
\cr
&+
\sum_{j=1}^{N} {\rm M}_{ij}^{\dag}(\theta) {\bm Q}_i(\theta) + \sum_{j=1}^{N} {\rm N}_{ji}^{\dag}(\theta) {\bm Q}_j(\theta) \Big) \cdot {\bm y}_i = 0.
\end{align}
Because this equation should hold for arbitrary ${\bm y}_i$, the phase sensitivity function ${\bm Q}_i(\theta)$ should satisfy the following set of adjoint equations:
\begin{align}
\omega \frac{d{\bm Q}_i(\theta)}{d\theta}
+ {\rm J}_i(\theta)^{\dag} {\bm Q}_i(\theta)
&+ \sum_{j=1}^{N} {\rm M}_{ij}^{\dag}(\theta) {\bm Q}_i(\theta)
\cr
&+ \sum_{j=1}^{N} {\rm N}_{ji}^{\dag}(\theta) {\bm Q}_j(\theta) = 0
\quad
\label{adjointapp}
\end{align}
for $i=1, 2, ..., N$. Finally, the normalization condition for ${\bm Q}_i(\theta)$ is obtained by differentiating Eq.~(\ref{phase1}) as
\begin{align}
\frac{d\theta}{dt} &= \sum_{i=1}^N \left. \frac{\partial \Theta}{\partial {\bm X}_i}\right|_{\{ {\bm X}_i = {\bm X}^{(0)}_i(\theta) \} } \cdot \frac{d {\bm X}_i^{(0)}(\theta)}{dt}
\cr
&= \sum_{i=1}^N {\bm Q}_i(\theta) \cdot \frac{d {\bm X}_i^{(0)}(\theta)}{dt} = \omega
\label{normalizationapp}
\end{align}
or
\begin{align}
\sum_{i=1}^N {\bm Q}_i(\theta) \cdot \frac{d {\bm X}_i^{(0)}(\theta)}{d\theta} = 1.
\label{normalizationapp2}
\end{align}

Thus, by calculating a $2\pi$-periodic solution to Eq.~(\ref{adjoint}) with the above normalization condition, we can obtain the phase sensitivity function ${\bm Q}_i(\theta)$ for each element $i$, characterizing the effect of tiny perturbations applied to the element $i$ when the phase of the whole network is $\theta$.
In actual numerical calculation, backward integration of Eq.~(\ref{adjoint}) with occasional normalization by Eq.~(\ref{normalization}) as proposed by Ermentrout~\cite{ermentrout10} is useful.

\subsection{Diffusively coupled oscillators on a network}

The following reaction-diffusion-type model on a network is often considered in the analysis of coupled oscillators on networks:
\begin{align}
\frac{d}{dt} {\bm X}_i(t) = {\bm F}_i({\bm X}_i) + {\rm D} \sum_{j=1}^N L_{ij} {\bm X}_j
\quad (i=1, 2, ..., N), 
\end{align}
where $L_{ij}$ is the $(i, j)$ component of ${N \times N}$ Laplacian matrix ${\rm L}$ of the network and ${\rm D}$ is a matrix of diffusion constants.
It is assumed that all elements share the same dimensionality $m$ and 
${\rm D} \in {\bm R}^{m \times m}$ is a square matrix.
The network is specified by an adjacency matrix ${\rm A} \in {\bm R}^{N \times N}$ of the network, whose $(i, j)$ component $A_{ij}$ is
$1$ when nodes $i$ and $j$ are connected and $0$ otherwise (generalization to weighted network is straightforward),
and the Laplacian matrix is defined as
\begin{align}
L_{ij} = A_{ij} - k_i \delta_{ij},
\end{align}
where $k_i = \sum_{j=1}^N A_{ij}$ is the degree of the network and $\delta_{ij}$ is the Kronecker's delta.

The coupling term in this case is given by
\begin{align}
{\bm G}_{ij}({\bm X}_i, {\bm X}_j) = {\rm D} ( L_{ij} {\bm X}_j ),
\end{align}
so that the Jacobian matrices ${\rm M}_{ij} \in {\bm R}^{m \times m}$ and ${\rm N}_{ij} \in {\bm R}^{m \times m}$ are given by
\begin{align}
{\rm M}_{ij} = 0, \quad {\rm N}_{ij} = {\rm D} L_{ij}.
\end{align}
The adjoint equations in this case are
\begin{align}
\omega \frac{d{\bm Q}_i(\theta)}{d\theta} + {\rm J}_i(\theta)^{\dag} {\bm Q}_i(\theta) + {\rm D}^{\dag} \sum_{j=1}^{N} L_{ji} {\bm Q}_j(\theta) = 0
\cr
(i=1, 2, ..., N),
\label{adjointlaplacian}
\end{align}
where ${\rm J}_i(\theta) \in {\bm R}^{m \times m}$ is the Jacobian matrix of ${\bm F}_i({\bm X}_i)$ at ${\bm X}_i = {\bm X}^{(0)}_i(\theta)$.

The above equations can be related to the adjoint partial differential equation for a spatially continuous reaction-diffusion system~\cite{nakao}
\begin{align}
\frac{\partial}{\partial t} {\bm X}({\bm r}, t) = {\bm F}({\bm X}({\bm r}, t), {\bm r}) + {\rm D} \nabla^2 {\bm X}({\bm r}, t)
\label{rdcontinuous}
\end{align}
exhibiting spatio-temporally rhythmic dynamics,
where ${\bm r} \in {\bm R}^d$ represents a position in $d$-dimensional continuous media, ${\bm X}({\bm r}, t) : {\bm R}^d \times {\bm R} \to {\bm R}^m$ is the $m$-component field variable at position ${\bm r}$ and time $t$, ${\bm F}({\bm X}, {\bm r}) \in {\bm R}^m$ describes the reaction dynamics at ${\bm r}$, and ${\rm D} \in {\bm R}^{m \times m}$ is a matrix of diffusion constants.

The set of adjoint equations (\ref{adjointlaplacian}) can be interpreted as a discretized generalization of the adjoint partial differential equation~\cite{nakao} for the phase sensitivity function ${\bm Q}({\bm r}, \theta )$ for a stable limit-cycle solution ${\bm X}^{(0)}({\bm r}, \theta)$ of Eq.~(\ref{rdcontinuous}),
\begin{align}
\omega \frac{\partial {\bm Q}({\bm r}, \theta)}{\partial \theta} + {\rm J}({\bm r}, \theta)^{\dag} {\bm Q}({\bm r}, \theta)
+ {\rm D}^{\dag} \nabla^2 {\bm Q}({\bm r}, \theta) = 0,
\end{align}
where ${\rm J}({\bm r}, \theta)$ is the Jacobian matrix of ${\bm F}({\bm X}, {\bm r})$ estimated at the state ${\bm X} = {\bm X}^{(0)}({\bm r}, \theta)$ and the position ${\bm r}$.
The normalization condition Eq.~(\ref{normalization}) can be also seen as a generalization for the continuous case,
\begin{align}
\int_V d{\bm r} \ {\bm Q}({\bm r}, \theta ) \cdot \frac{\partial {\bm X}^{(0)}({\bm r}, \theta)}{\partial \theta} = 1,
\end{align}
where $V$ is the considered domain.
Formal correspondence between the adjoint equations for the network and for the continuous media is apparent, where the index $i$ corresponds to the position ${\bm r}$ and the Laplacian matrix $L_{ij}$ corresponds to the Laplacian operator $\nabla^2$.

\subsection{Coupling matrix and collective dynamics of the network}

The following coupling matrix, whose components are randomly and independently drawn from a uniform distribution $[-0.6, 0.6]$, is used throughout numerical simulations.
With this coupling matrix and the parameters of the elements given in Sec. III-A (\#1-\#7: excitable, \#8-\#10: oscillatory), the network started from a uniform initial condition, $u_i=1$ and $v_i=1$ for all $i=1, 2, ..., 10$, converges to a limit-cycle attractor of period $T \simeq 75.73$ in the $20$-dimensional state space, which corresponds to the collectively oscillating state of the network.
Despite high-dimensionality of the network and random coupling between the elements, this limit-cycle attractor is robust and the network always converged to this attractor even if the network was started from $1000$ different random initial conditions (initial values of $u_i$ and $v_i$ randomly and independently chosen from a uniform distribution $[-10, 10]$).
This particular limit-cycle solution is used for all numerical simulations in the example.

\begin{widetext}
\begin{align}
{\rm K} =
\begin{pmatrix}
0.000 & 0.409 & -0.176 & -0.064 & -0.218 & 0.464 & -0.581 & 0.101 & -0.409 & -0.140 \\
0.229 & 0.000 & 0.480 & -0.404 & -0.409 & 0.040 & 0.125 & 0.099 & -0.276 & -0.131 \\
-0.248 & 0.291 & 0.000 & -0.509 & -0.114 & 0.429 & 0.530 & 0.195 & 0.416 & -0.597 \\
-0.045 & 0.039 & 0.345 & 0.000 & 0.579 & -0.232 & 0.121 & 0.130 & -0.345 & 0.463 \\
-0.234 & -0.418 & -0.195 & -0.135 & 0.000 & 0.304 & 0.124 & 0.038 & -0.049 & 0.183 \\
-0.207 & 0.536 & -0.158 & 0.533 & -0.591 & 0.000 & -0.273 & -0.571 & 0.110 & -0.354 \\
0.453 & -0.529 & -0.287 & -0.237 & 0.470 & -0.002 & 0.000 & -0.256 & 0.438 & 0.211 \\
-0.050 & 0.552 & 0.330 & -0.148 & -0.326 & -0.175 & -0.240 & 0.000 & 0.263 & 0.079 \\
0.389 & -0.131 & 0.383 & 0.413 & -0.383 & 0.532 & -0.090 & 0.025 & 0.000 & 0.496 \\
0.459 & 0.314 & -0.121 & 0.226 & 0.314 & -0.114 & -0.450 & -0.018 & -0.333 & 0.000 \\
\end{pmatrix}
\label{actualK}
\end{align}
\end{widetext}

Detailed characterization of the collective dynamics that can take place in general networks of randomly coupled oscillatory and excitable FitzHugh-Nagumo elements is a difficult task and is not the focus of the present study.
Here, we only briefly describe numerical results for the network of $N=10$ FitzHugh-Nagumo elements with the coupling matrix ${\rm K}$ whose elements were drawn independently from uniformly distributed random variables as described above.
The following qualitative characteristics were common to several different realizations of the random matrix ${\rm K}$ with the same statistics.

Firstly, when the overall coupling intensity of the network was varied by using $c K_{ij}$ in Eq.~(\ref{fhncoupling}) instead of $K_{ij}$, where the parameter $c > 0$ was used to control the overall coupling intensity, the network exhibited chaotic dynamics for small $c$ (roughly $c < 0.2$ for the above ${\rm K}$), stable limit-cycle dynamics for intermediate values of $c$ ($0.2 < c < 1.4$), and stable fixed point for large $c$ ($c > 1.4$). In between the chaotic and oscillatory regimes, narrow regimes with quasi-periodic dynamics were also observed.
Secondly, qualitative behavior of the network did not change largely even if the number of oscillatory elements was varied between $1$ and $9$ when $c=1$.
In a few cases, the network could possess two coexisting limit-cycle attractors, and the network started from random initial conditions converged to either of those attractors. These coexisting limit-cycle attractors had similar but slightly different periods and individual trajectories of the elements.
In contrast, when all elements of the network were oscillatory, the collective oscillation was qualitatively different from the other cases with excitable elements and the network possessed many coexisting limit-cycle attractors. These attractors also had similar but slightly different periods and individual trajectories.
Finally, when all the elements were excitable, no collective oscillation was observed when the network started from a uniform initial condition.

These numerical results suggest that the collectively oscillating solution used as an example in the present study is typical and robust, though, of course, the above is only a brief numerical survey of the network of randomly coupled FitzHugh-Nagumo elements used in this study and much more detailed analysis is necessary to fully characterize general dynamical properties of such networks.
Note also that the phase reduction theory developed in the present study is applicable to any stable limit-cycle attractor of an arbitrary network of coupled dynamical elements given by Eq.~(\ref{1}), provided that the perturbation (e.g. mutual coupling) applied to the network is sufficiently weak.


\begin{thebibliography}{99}

\bibitem{winfree80}
  A.~T.~Winfree,
  {\it The Geometry of Biological Time}
  (Springer, New York, 1980; Springer, Second Edition, New York, 2001).

\bibitem{pikovsky01}
  A.~Pikovsky, M.~Rosenblum, and J.~Kurths,
  {\it Synchronization: A Universal Concept in Nonlinear Sciences}
  (Cambridge University Press, Cambridge, 2001).
  
\bibitem{strogatz03}
  S.~H.~Strogatz,
  {\it Sync: How Order Emerges from Chaos in the Universe, Nature, and Daily Life}
  (Hyperion Books, New York, 2003).

\bibitem{hudson0}
I.~Z.~Kiss, W.~Wang, J.~L.~Hudson,
Experiments on arrays of globally coupled periodic electrochemical oscillators,
J. Phys. Chem. B {\bf 103} 11433 (1999).

\bibitem{hudson1}
W.~Wang, I.~Z.~Kiss, J.~L.~Hudson,
Experiments on arrays of globally coupled chaotic electrochemical oscillators: Synchronization and clustering,
Chaos {\bf 10} 248 (2000).

\bibitem{hudson2}
I.~Z.~Kiss, Y.~Zhai, and J.~L.~Hudson,
Emerging Coherence in a Population of Chemical Oscillators,
Science {\bf 296}, 1676 (2002).

\bibitem{hudson3}
I.~Z.~Kiss, Y.~Zhai, and J.~L.~Hudson,
Predicting Mutual Entrainment of Oscillators with Experiment-Based Phase Models,
Phys. Rev. Lett. {\bf 94}, 248301 (2005).

\bibitem{hudson4}
I.~Z.~Kiss, Y.~Zhai, and J.~L.~Hudson,
Characteristics of Cluster Formation in a Population of Globally Coupled Electrochemical Oscillators: An Experiment-Based Phase Model Approach,
Prog. Theoret. Phys. Suppl. {\bf 161}, 99 (2006).

\bibitem{hudson5}
I.~Z.~Kiss, C.~G.~Rusin, H.~Kori, J.~L.~Hudson,
Engineering complex dynamical structures: Sequential patterns and desynchronization, 
Science {\bf 316} 1886 (2007).

\bibitem{hudson6}
H.~Kori, C.~G.~Rusin, I.~Z.~Kiss, J.~L.~Hudson,
Synchronization engineering: Theoretical framework and application to dynamical clustering,
Chaos {\bf 18} 026111 (2008).

\bibitem{motter13}
A.~E.~Motter, S.~A.~Myers, M.~Anghel, and T.~Nishikawa,
Spontaneous synchrony in power-grid networks,
Nature Phys. {\bf 9}, 191 (2013).

\bibitem{kuramoto}
  Y.~Kuramoto,
  {\it Chemical Oscillations, Waves, and Turbulence}
  (Springer, New York, 1984; Dover, New York, 2003).
  
\bibitem{hoppensteadt97}
  F.~C.~Hoppensteadt and E.~M.~Izhikevich,
  {\it Weakly Connected Neural Networks}
  (Springer, New York, 1997).
  
\bibitem{ermentrout10}
  G.~B.~Ermentrout and D.~H.~Terman,
  {\it Mathematical Foundations of Neuroscience}
  (Springer, New York, 2010).

\bibitem{brown04}
  E.~Brown, J.~Moehlis, and P.~Holmes,
  On the phase reduction and response dynamics of neural oscillator populations,
  Neural Comput. {\bf 16}, 673 (2004).

\bibitem{ashwin16}
  P.~Ashwin, S.~Coombes, and R.~Nicks,
  Mathematical frameworks for oscillatory network dynamics in neuroscience,
  J. Math. Neurosci. {\bf 6}, 1 (2016).

\bibitem{nakao16}
  H.~Nakao,
  Phase reduction approach to synchronization of nonlinear oscillators,
  Contemp. Phys. {\bf 57}, 188 (2016).
  

\bibitem{kawamura1}
  Y.~Kawamura, H.~Nakao, K.~Arai, H.~Kori, and Y.~Kuramoto,
  Collective phase sensitivity,
  Phys. Rev. Lett. {\bf 101}, 024101 (2008).

\bibitem{kori}
	H.~Kori, Y.~Kawamura, H.~Nakao, K.~Arai, and Y.~Kuramoto,
	Collective dynamical response of coupled oscillators with general network structure,
	Phys. Rev. E {\bf 80}, 036207 (2009).

\bibitem{kawamura2}
	Y.~Kawamura, H.~Nakao, and Y.~Kuramoto,
	Collective phase description of globally coupled excitable elements,
	Phys. Rev. E {\bf 84}, 046211 (2011).

\bibitem{kawamura3}
	Y.~Kawamura, H.~Nakao, K.~Arai, H.~Kori, and Y.~Kuramoto,
	Phase synchronization between collective rhythms of globally coupled oscillator groups: Noisy identical case,
	Chaos {\bf 20}, 043109 (2010).

\bibitem{kawamura4}
	Y.~Kawamura, H.~Nakao, K.~Arai, H.~Kori, and Y.~Kuramoto,
	Phase synchronization between collective rhythms of globally coupled oscillator groups: Noiseless non-identical case,
	Chaos {\bf 20}, 043110 (2010).

\bibitem{kawamura44}
	Y.~Kawamura,
	Phase synchronization between collective rhythms of fully locked oscillator groups,
	Sci. Rep. {\bf 4}, 4832 (2014).

\bibitem{kawamura5} 
  Y.~Kawamura and H.~Nakao,
  Collective phase description of oscillatory convection,
  Chaos {\bf 23}, 043129 (2013).

\bibitem{kawamura6}
  Y.~Kawamura and H.~Nakao,
  Phase description of oscillatory convection with a spatially translational mode,
  Physica D {\bf 295-296}, 11 (2015).

\bibitem{nakao}
	H.~Nakao, T.~Yanagita, and Y.~Kawamura,
	Phase-reduction approach to synchronization of spatiotemporal rhythms in reaction-diffusion systems,
	Phys. Rev. X {\bf 4}, 021032 (2014).
 

\bibitem{moehlis06}
  J.~Moehlis, E.~Shea-Brown, and H.~Rabitz,
  Optimal inputs for phase models of spiking neurons,
  J. Comput. Nonlin. Dyn. {\bf 1}, 358 (2006).
  
\bibitem{harada10}
  T.~Harada, H.-A.~Tanaka, M.~J.~Hankins, and I.~Z.~Kiss,
  Optimal waveform for the entrainment of a weakly forced oscillator,
  Phys. Rev. Lett. {\bf 105}, 088301 (2010).
  
\bibitem{zlotnik13}
  A.~Zlotnik, Y.~Chen, I.~Z.~Kiss, H.~Tanaka, and J.-S.~Li,
  Optimal waveform for fast entrainment of weakly forced nonlinear oscillators,
  Phys. Rev. Lett. {\bf 111}, 024102 (2013).
  
\bibitem{pikovsky15}
   A.~Pikovsky,
   Maximizing coherence of oscillations by external locking,
   Phys. Rev. Lett. {\bf 115}, 070602 (2015).
         
\bibitem{tanaka15}
  H.-A.~Tanaka, I.~Nishikawa, J.~Kurths, Y.~Chen, and I.~Z.~Kiss,
  Optimal synchronization of oscillatory chemical reactions
  with complex pulse, square, and smooth waveforms signals maximizes Tsallis entropy,
  Europhys. Lett. {\bf 111}, 50007 (2015).
  
\bibitem{zlotnik16}
  A.~Zlotnik, R.~Nagao, I.~Z.~Kiss, and J.-S.~Li,
  Phase-selective entrainment of nonlinear oscillator ensembles,
  Nature Comm. {\bf 7}, 10788 (2016).

\bibitem{kawamura16}
	Y. Kawamura and H. Nakao,
	Optimization of noise-induced synchronization of oscillator networks,
	Phys. Rev. E {\bf 94}, 032201 (2016).

\bibitem{shirasaka17}
	S.~Shirasaka, N.~Watanabe, Y.~Kawamura, and H.~Nakao,
	Optimizing stability of mutual synchronization between a pair of limit-cycle oscillators with weak cross coupling,
	Phys. Rev. E {\bf 96}, 012223 (2017).
	
\bibitem{kawamura17}
	Y.~Kawamura, S.~Shirasaka, T.~Yanagita, and H.~Nakao,
	Optimizing mutual synchronization of rhythmic spatiotemporal patterns in reaction-diffusion systems,
	Physical Review E {\bf 96}, 012224 (2017).

\end{thebibliography}
\end{document}